\documentclass[article,twocolumn, reprint,nofootinbib, amssymb]{revtex4-1}

\usepackage{graphicx}
\usepackage{color}
\usepackage{amssymb}  
\usepackage{amsthm}
\usepackage{amsmath}

\begin{document}

\title{ Cooling of Neutron Stars admixed with light dark matter: a case study }

\author{M. Ángeles Pérez-García $^{1}$, H. Grigorian$^{2,3,4}$, C. Albertus$^{1}$, D. Barba  $^{1}$ and J. Silk $^{5,6,7}$}

\affiliation{$^{1}$ Department of Fundamental Physics, Universidad de Salamanca, Plaza de la Merced S/N E-37008, Salamanca, Spain\\
$^2$ Laboratory of Information Technologies, JINR Dubna, Dubna, Russia\\
$^3$  Computational Physics and IT Division, A.I. Alikhanyan National Science Laboratory, Armenia\\
$^4$  Department for Theoretical Physics, Yerevan State University, Yerevan, Armenia\\
 $^5$  Institut d'Astrophysique,  UMR 7095 CNRS, Universit\'e Pierre et Marie Curie, 98bis Blvd Arago, 75014 Paris, France\\ $^6$  Department of Physics and Astronomy, The Johns Hopkins University, Homewood Campus, Baltimore MD 21218, USA\\
$^7$ Beecroft Institute of Particle Astrophysics and Cosmology, Department of Physics, University of Oxford, Oxford OX1 3RH, UK,\\
}%

\date{\today}

\begin{abstract}
 Neutron Stars (NSs) are born as hot, lepton-rich objects that evolve according to the standard paradigm through subsequent stages where they radiate the excess of energy by emitting, first, neutrinos and, later on, photons. Current descriptions based on Standard Model calculations cannot fully explain all the existing cooling data series for the dozens of objects that have been reported. In this work, we consider the intriguing possibility that cooling NSs could be actually  admixed with a fraction of light dark matter (LDM), $\chi$. We focus on a particular case study assuming a generic light candidate with mass  $m_\chi=0.1$ $\rm GeV/c^2$ that undergoes self-annihilating reactions through pseudoscalar mediators producing neutrinos in the final state. We include one  additional feature, allowing thermal conduction from LDM while inside the dark core. By performing simulations of the temperature evolution in the NS, we find that cooling patterns could  be distorted by the presence of LDM and discuss these results in light of their observability.
\end{abstract}

\maketitle

\section{Introduction}

Dark Matter (DM) remains  one of the fundamental challenges in modern Cosmology and, in general, to the current status of modern Physics. Different  accumulated evidence over decades points towards this dark component \cite{dark}. DM is a crucial ingredient, necessary in order to produce the large inhomogeneities
that form structures in the Universe today. Indeed, looking at the Cosmic Microwave Background, we measure very small
temperature fluctuations, proportional to the ratio of density fluctuations in the baryonic
density and its background value. The potential problem of the growth of overdensities in the matter-dominated epoch in the Universe is under control, provided a new {\it dark} species, which decoupled from photons much earlier than the baryons did, eventually
makes structure formation a consistent possibility.
In most of the currently envisioned scenarios, DM is under the form of a new, yet undetected, particle that should be produced in an extension of the Standard Model (SM) of particle physics.  This massive particle, $\chi$,  produced in the early Universe when the temperature, $T$, fell below the DM mass, $m_\chi$, allowed the
equilibrium distribution to become Boltzmann-suppressed, with a factor given by $\sim \mathrm{exp}(-m_{\chi}/T)$, where we use $c=k_B=1$. At some point, the expansion rate became larger than the thermally-averaged self-annihilation rate, freezing-out the relative density of DM particles. The observed relic abundance of DM points to values of the thermally-averaged self-annihilation rate $\left<\sigma v\right> \sim 3\times 10^{-26} \,\rm cm^3 s^{-1}$. This is typical of  weakly-interacting particles and is one of the key parameters in the search for indirect probes of the existence of DM \cite{iocco}, in particular of the self-annihilation of DM into SM particles. 

Light dark matter (LDM) with a mass in the keV to GeV range coupling to SM particles via a new force mediator provides
a well-motivated alternative to the WIMP hypothesis \cite{new}.
The current status on the constraints on LDM are weaker as compared to more massive candidates due to lower sensitivity of direct detection searches. In this sense sub-GeV DM scattering with nucleons ($N$) is constrained to cross sections less than  $\sim 10^{-31}$ $\rm cm^2$ \cite{supercdms}, thus leaving  LDM candidates poorly explored by direct and also indirect searches \cite{coogan}. However the situation is slightly improved when scattering on electrons is considered achieving  sensitivities of $\sim 10^{-38}$ $\rm cm^2$ \cite{sigmae1, sigmae2,sigmae3,sigmae4,sigmae5}.

A number of works have previously explored NS stability including several types of DM candidates e.g. free and interacting Fermi gas \cite{sch}, non self-annihilating self-interacting matter \cite{tolos} or milli-charged DM \cite{kouvper}. Another popular model, that  we will adopt in this work, is that where it is assumed that DM undergoes self-annihilating reactions inside a dark core. For this latter case the presence of DM has been indirectly searched for via modifications of standard luminosities or spectra in stars \cite{wd}.

 Regarding NSs, current X-ray and multiband missions have been able to measure cooling curves i.e. isolated pulsar surface temperatures  $T_{\mathrm{s}}^{\infty}(t)$ or luminosities $L_{\gamma}^{\infty}(t)$ with increasing levels of precision. Generally speaking, one can distinguish three main cooling stages \cite{cool}. An  initial thermal relaxation stage lasts up to $\sim100$ yr when the surface temperature reflects the thermal state of the crust. Later, a neutrino cooling stage occurs where the main contribution comes from the neutrino luminosity $L_{\nu}$ from reactions involving core nucleons during $t \lesssim 10^{5}$ yr. Finally, a photon cooling stage where photons dominate and $L_{\nu} \ll L_{\gamma}$ for  $t \gtrsim 10^{5}\, \mathrm{yr}$. The star cools via photon emission from the surface.
In general, in the standard minimal cooling scenario, heating sources are neglected although some authors have discussed mechanisms of this kind such as frictional dissipation of rotational energy, deviations from beta equilibrium or ohmic dissipation \cite{umeda,reis}. There is also an additional mechanism based in the gapless color-flavor locked quark matter as described in \cite{alford}. In that work the authors show that the gCFL phase provides heat capacities and neutrino luminosities capable of keeping the NS warm at ages of $10^8$ years displaying surface temperatures $\gtrsim 10^4 $ K as long as there is an inner quark core sized $\sim 3$ km.

\section{Dark matter inside the cooling Neutron Star}

To model the cooling curve for a NS, we use the set of  general relativistic equations of thermal evolution for a spherically symmetric star under the form \cite{thorne},

\begin{equation}
\frac{\mathrm{e}^{-\lambda-2 \Phi}}{4 \pi r^{2}} \frac{\partial}{\partial r}\left(\mathrm{e}^{2 \Phi} L\right)=-Q+Q_{\mathrm{h}}-\frac{c_{V}}{\mathrm{e}^{\Phi}} \frac{\partial T}{\partial t},
\label{eq1}
\end{equation}

\begin{equation}
\frac{L}{4 \pi \kappa r^{2}}=\mathrm{e}^{-\lambda-\Phi} \frac{\partial}{\partial r}\left(T \mathrm{e}^{\Phi}\right)
\label{eq2}
\end{equation}
    
where $r$ is the radial coordinate, $Q$ is the neutrino emissivity, $c_{V}$ is the heat capacity per unit volume, $\kappa$ is the thermal conductivity, and $L$ is the local luminosity. $\Phi(r)$ and $\lambda(r)$ are the metric functions. The function $\Phi(r)$ specifies the gravitational redshift, while $\lambda(r)$ describes the gravitational distortion of radial scales, $\mathrm{e}^{-\lambda}=\sqrt{1-2 G m(r) /r}$, where $m(r)$ is the gravitational mass enclosed within a sphere of radius $r$. At the stellar surface, $\Phi(R)=-\lambda(R)$. For completeness, we have introduced $Q_{\mathrm{h}}\equiv Q_{\chi \chi}$, which represents the rate of energy production (heating) by LDM annihilation into final neutrino states that will be detailed later.
In the outermost stellar layers, thermal conduction is radiative, however deeper in the crust, thermal conductivity is provided by electrons  while in the core it is produced by other fermions ($f$), e.g. electrons, nucleons. Note that for a set of  additional heat carriers it attains a value $\kappa=\sum_{i=f,\chi} \kappa_i$.

We write $L_{\gamma}=4 \pi R^{2} \sigma_B T_{\mathrm{s}}^{4}$ as the thermal photon luminosity in the local reference frame of the star, being  $T_{\mathrm{s}}$ the effective temperature and $\sigma_B$ is the Stefan-Boltzmann constant. The apparent (redshifted) effective temperature $T_{\mathrm{s}}^{\infty}$ and luminosity $L_{\gamma}^{\infty}$, as detected by a distant observer, are
$T_{\mathrm{s}}^{\infty}=T_{\mathrm{s}} \sqrt{1-r_{g} / R}$ and $L_{\gamma}^{\infty}=L_{\gamma}\left(1-r_{g} / R\right)$ where $r_{g}=2 G M \approx 2.95 M / M_{\odot} \,\mathrm{km}$ is the Schwarzschild radius. One often introduces the apparent radius $R_{\infty}=R / \sqrt{1-r_{g} / R}$ which an observer would measure.

In this scenario, the stellar progenitor and the actual NS have been accumulating DM over time due to gravitational capture. The efficiency of the different processes producing an admixed star has already been explained in the literature \cite{cermeno,thermaliz}. Note that in order for the NS to be opaque to the LDM candidate under inspection, the latter should have a scattering 
cross-section with nucleons (the main ingredient inside the dense stellar core) larger than a critical value $\sigma_0\sim 10^{-46} \,\rm cm^2$, which is fulfilled in our case study. Besides, evaporation is negligible. From this, the thermalized local LDM distribution has a radius 
$R_{\chi} \sim 0.8 \,\mathrm{km} \left(\frac{0.1\, \rm GeV} {m_\chi} \frac{2 \rho_{0}} {\rho_{N}}\frac{T}{0.1\,\rm MeV}\right)^{1/2}$, 
much smaller than the stellar (core) radius $R_\chi \ll R\lesssim 14$ km. In the previous we have assumed that the inner km-sized LDM core is spread over an approximately constant baryonic density region with value $\rho_{N}$ relevant for the NS mass range we will explore $M \in [1,1.9]M_\odot$. $\rho_0\sim 2.4\times 10^{14}\,\rm g/cm^3$ is the nuclear saturation density. 

In order to expose the differences in the NS cooling patterns  arising from the specific choice of our dark candidate inside the core, we will analyze in detail two LDM features: self-annihilation and thermal conduction. The theoretical framework of energy transport by conduction of weakly interacting DM was first studied in a solar environment \cite{conduc,gouldraffelt}. More refined treatments have been developed, see e.g. \cite{vincent}. As we will explain, the energy transport and heat flow in the NS will be affected by these extra contributions from the dark components.

The amount of LDM captured by the opaque star \cite{cermeno1}  is an increasing function of time as the $\chi$ population sinks towards the center \cite{gould,capture}. However as the local DM density grows, the self-annihilation reactions will start to be efficient leading to potentially  dramatic changes \cite{herrero}. In more detail, the dynamical number of LDM particles,  $N_{\chi}$, is obtained by solving 
\begin{equation}
N_\chi (t)\simeq N_{\chi,0}+ \frac{dN_\chi}{dt} (t-t_0),
\end{equation}
where the ratio of change in $N_{\chi}$, can be obtained as a function of time from an initial population $N_{\chi, 0}$ at time $t_0$. In a simplified scheme, the dynamical equation for the DM population can be written as the result of two processes that will compete among each other, i.e. capture and self-annihilation inside the dark core
\begin{equation}
\frac{d N_{\chi}}{d t}=C_{\chi}-C_{a} N_{\chi}^{2}.
\end{equation}
The DM capture rate can be approximated by 
\small
\begin{equation}
C_{\chi}\simeq 5.6 \times 10^{26}\left(\frac{M}{1.5 M_{\odot}}\right)\left(\frac{R}{14 \mathrm{~km}}\right)\left(\frac{0.1 \mathrm{GeV}}{m_{\chi}}\right)\left(\frac{\rho_{\chi}}{0.4 \frac{\mathrm{GeV}}{\mathrm{cm}^{3}}}\right) \,\mathrm{s}^{-1},
\end{equation}
\normalsize
whereas the self-annihilation rate is $C_{a}\simeq \frac{\left<\sigma v\right>}{R_\chi^3}$ with the thermally-averaged self-annihilation rate $\left<\sigma v\right> \sim 3\times 10^{-26} \,\rm cm^3 s^{-1}$,
\begin{equation}
C_a\simeq 2\times 10^{-42} \left(\frac{0.1\, \,\rm GeV} {m_\chi} \frac{2 \rho_{0}} {\rho_{N}}\frac{T}{0.5\,\rm MeV}\right)^{-3/2}\,\rm s^{-1}.
\end{equation}

Besides, at initial time $t_0$, the dark population inside the NS is
\begin{equation}
N_{\chi, 0}=1.5 \times 10^{39}\left(\frac{\rho_{\chi}}{0.4 \frac{\mathrm{GeV}}{\mathrm{cm}^{3}}}\right)\left(\frac{0.1\, \mathrm{GeV}}{m_{\chi}}\right)\left(\frac{\sigma_{s}}{10^{-44} \mathrm{~cm}^{2}}\right),
\end{equation}
where $\sigma_{s} \equiv \sigma_{\chi-N}$ is the $\chi-N$ scattering cross-section that we will take in this work as $\sigma_{s} \in\left[10^{-46}-10^{-31}\right] \mathrm{cm}^{2}$ using current experimental constraints as discussed in the Introduction section. The average DM density in the solar neighbourhood is $\rho_{\chi,0}\sim 0.4$  $\rm GeV/cm^3$. However it is expected that inside some globular clusters DM can be enhanced by up to several orders of magnitude, e.g. NGC 6266, 47 Tuc \cite{brown}, as is the case for the faintest dwarf galaxies.

Inside the NS, we focus on the reactions taking place and model them with a pseudoscalar mediator $a$ with $m_\chi \in [0.1,30]$ GeV $(m_a \in[0.05,1]\,\rm GeV)$, see our previous work \cite{lineros}. The self-annihilations of DM proceed into two-body fermionic states $(f), \chi \chi \rightarrow f \bar{f}$ or two pseudoscalar boson states,  $\chi \chi \rightarrow a a$ with subsequent decay $a \rightarrow f \bar{f}$. Further, we will have  interest in $f=\nu$, the neutrino channel in our cooling scenario.
 
 One of the key quantities that dictate NS internal stellar energetic balance is the local energy emissivity, $Q_{E}=\frac{d E}{d V d t}$ (energy produced per unit volume per unit time). Formally, the expression for $Q_{E}$ is given by 
\begin{equation}
Q_{E}=4 \int d \Phi\left(E_{1}+E_{2}\right)|\bar{\mathcal{M}}|^{2} \mathcal{F},
\end{equation}

where $E_1+E_2$ is the energy carried by neutrinos in the reaction $1+2\rightarrow 3+4$ with matrix element $\mathcal{M}$. Particularizing to the heating processes related to DM $Q_h$ in Eq. ({\ref{eq1}) the global phase space factor is given by $\mathcal{F}=f_{\chi}\left(E_{1}\right)f_{\chi}\left(E_{2}\right) \left(1-f_{f}\left(E_{3}\right)\right)\left(1-f_{f}\left(E_{4}\right)\right)$. $f_{\chi}, f_{f}$ are the local stellar distribution functions for LDM and fermionic particles, respectively, containing density and temperature dependence. For the sake of simplicity we will consider the reasonable assumption that DM follows a non-relativistic Maxwell-Boltzmann distribution inside the core when thermalized and neutrinos produced will be,  in a continuous steady state of annihilations once stabilized in the dark core from the gravitational capture of DM by the NS.

When modeling the dark contribution to the cooling NS from this type of models we found that the emissivity is a function of the  temperature $T$, for a specific realization. For the test cases, we will analyze the dark emissivity cast into the approximate  expression in the central km as (see Fig. 2 in \cite{lineros})
\begin{equation}
 Q_h\equiv Q_{\chi \chi}(T)={10^{\alpha}}\left(\frac{T}{0.1\, \rm MeV}\right)^{-3} \,\rm erg \,cm^{-3} \,s^{-1}, 
\label{qxx}
\end{equation}

with $19 < \alpha < 21$. Note the sign difference appearing in Eq. \eqref{eq1}  when compared  to a regular cooling process $Q$ as the self-annihilation reactions contribute to the heating in the inner dark core and decrease with increasing temperature as the thermal volume grows. It is well known that standard direct URCA or modified URCA (MURCA) cooling processes provide emissivities $Q_{E}^{\mathrm{URCA}} \sim 10^{27} \mathcal{R}\left(\frac{T}{0.1 \,\mathrm{MeV}}\right)^{6}$ erg\, $\mathrm{cm}^{-3} \mathrm{~s}^{-1}$ and $Q_{E}^{\mathrm{MURCA}} \sim 10^{21} \mathcal{R}\left(\frac{T}{0.1\, \mathrm{MeV}}\right)^{8} \mathrm{erg}\, \mathrm{cm}^{-3} \mathrm{~s}^{-1}$, respectively. Higher central densities are needed to activate the URCA reactions and can only be  realized in the most massive NSs. Typical energetic scales can be obtained from the conversion factor 1 $\mathrm{MeV} \sim 10^{10}\, \mathrm{~K}$. $\mathcal{R}$ is a reduction function of order unity describing the superfluid effects in the neutron and proton branches of those reactions involving nucleons \cite{page}. Note that for temperatures $T\lesssim 10^9$ K the MURCA and $\chi$ self-annihilation attain similar strengths. As we will explain, it could happen that the efficiency of this annihilating DM mechanism could act on a temporary basis, due to the dynamics of the aforementioned competing effects allowing for a given finite time $\tau_{\chi}$ where it is active. 
\begin{figure}[t]
\begin{center}
\includegraphics [width=1.3\columnwidth, angle=0] {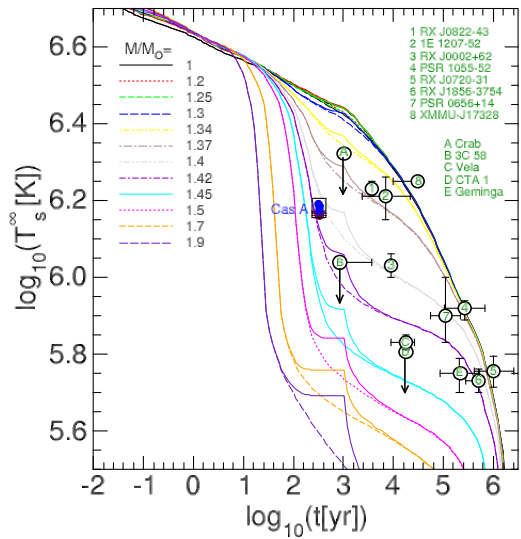}
\caption{Surface temperature as a function of time for NSs with masses $M\in[1,1.9] M_\odot$ with the effect of self-annihilating LDM ($m_\chi=0.1$ GeV) originating a plateau or without LDM (continuous decline). Existing series of cooling NSs are plot as well. See text for details.}
\label{fig1}
\end{center}
\end{figure}

We will include another feature in the LDM sector by considering that may act as a heat-conducting agent. Since LDM interaction with ordinary matter is weak, it could be an efficient energy carrier redistributing heat in the thermal volume where it is mostly concentrated. This dark conductivity will add to the already existing lepton and hadronic standard components that do transport heat in the NS \cite{page}. The conductivity for the LDM distribution in the inner dark core can be written into the approximate form as $\kappa_\chi\sim v_\chi \lambda_\chi n_\chi$ where $v_\chi=\sqrt{3 T/m_\chi}$ is the thermalized velocity and the typical size of $\chi-N$ mean free path is $\lambda_\chi\sim {(\sigma_{\chi N} n_N)}^{-1}$. Note however that in practice  $\lambda_\chi$  associated to the massive $\chi$ will be at most a few times that of the $R_\chi$ value since the gravitational potential will exert its pull towards the core of the star.
Thus LDM involves an additional contribution to the standard cooling fermionic luminosity $\kappa=\sum_f \kappa_f+\kappa_\chi$.

In this picture, we expect the larger amount of effective dark carriers in the star will be able to speed up internal transport in the NS inner core. 

\section{Results}

In this section we analyze our results. We have solved the set of equations \eqref{eq1} and \eqref{eq2} using the prescribed $Q_h$ as obtained from annihilating DM models A, B, C in \cite{lineros} and the nuclear equation of state HDD as given in \cite{eos} including proton superconductivity with the parametrization given by EEHOr (no pion condensation) \cite{hoviksc}.
In Figure~\ref{fig1} we  plot the surface temperature as a function of time for NSs with masses $M\in[1,1.9] M_\odot$ with the effect of self-annihilating conducting DM for $m_\chi=0.1$ GeV. Experimental temperature constrains are also shown for some cooling NSs measured in the literature. We can clearly see a plateau signaling the existence of a dramatic halt in the rapid cooling behaviour originating earlier and lasting up to  $\tau_\chi \sim 10^3$ yr (solid lines) for the most massive NSs. Due to the energy injection the DM self-annihilation can temporarily stop the decline in temperature. Later on, it is no longer efficient and ordinary cooling due to neutrinos and photons proceeds, decreasing the temperature again, but, importantly, shifting the initial drop departure point. Similar average effects were found for global emissivities, resulting in much fainter objects \cite{kouvaris}. Other quoted mechanisms, such as the gCFL phase \cite{alford} can warm an old NS provided a few km-sized quark core is present or dark kinetic heating \cite{acevedo}. This fact could be experimentally accessible using the infrared telescopes operating in the coming future such as James Webb Space Telescope \cite{jwt} or the Thirty Meter Telescope \cite{tmt}.

\begin{figure}[h!]
\begin{center}
\includegraphics [width=1.2\columnwidth, angle=0] {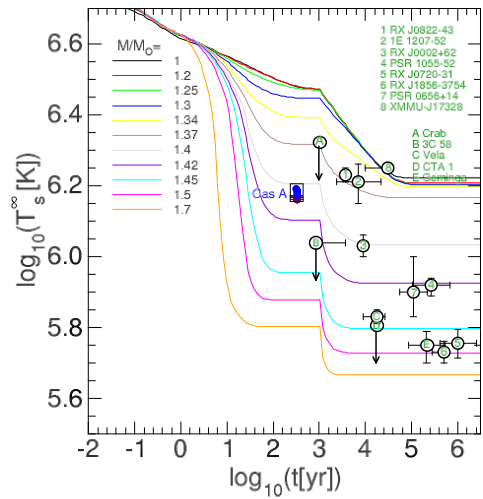}
\caption{Surface temperature as a function of NS age with masses $M\in[1,1.7] M_\odot$ including self-annihilating conducting DM ($m_\chi=0.1$ GeV). $\chi$ emissivity has been enhanced a factor 5 larger than in Figure~\ref{fig1}.  LDM enhanced processes are active up to $\tau\sim 10^3$ yr, followed by a period of decline, and again for $t\gtrsim 1.5 \times 10^3$ yr. See text for details.}
\label{fig2}
\end{center}
\end{figure}

In Figure~\ref{fig2}, we plot surface temperature as a function of NS age with masses $M\in[1,1.7]M_\odot$ including self-annihilating and conducting LDM ($m_\chi=0.1$ GeV). $\chi$ emissivity has been enhanced a factor 5 larger than that in Figure~\ref{fig1}.  $\chi$ processes are active during $\tau\sim 10^3$ yr, followed by a period of interruption, and active again for  $t\gtrsim 1.5 \times 10^3$ yr. Remarkably, some observations, see for example RX J7020-31, RX J1856-3754, Geminga  could be interpreted as cooling NSs with no LDM and $M \in [1,1.42]M_\odot$ (see  Figure~\ref{fig1}) or having a similar larger mass $M=1.5M_\odot$ in the interrupted LDM heating picture. Complementary refined mass measurements seem necessary to possibly help disentangle the actual scenario. Although not shown in our plot, the recent measurement of the coldest NS to date,  PSRJ2144-3933 fitted as a $3 \times 10^8$ yr object with a mass 1.4$M_{\odot}$ and unredshifted temperature $T<42000$ K puts some constraint on the duration $\tau_\chi$ and stregth of the self-annhilation processes. Note that in Figure~\ref{fig1} at those ages $T$ remains unchanged with respect to DM-free case for the case under study. However, for the scenario in Figure~\ref{fig2}, matching luminosities must drop at least 1.5 orders of magnitude in redshifted temperature and this advises that emissivities, proportional to $m^{5/2}_\chi$  should constraint DM masses at the $\sim$10 MeV range. Such low $\chi$ masses should be carefully considered in the stellar population as the low-energy modes from lepton dynamics and collective modes will add complexity to the overall picture.  Since we do not follow our simulations to such long ages this aspect remains for future work.

%
\section{Conclusions}

In this contribution, we have analyzed the cooling of NSs with a light dark mater core acquired by gravitational capture. We have considered a pseudoscalar-mediated interaction model where $\chi$ self-annihilation reactions producing final state neutrinos can be treated as a source of heating inside the star through an emissivity term  ($Q_h$) injecting energy in the core along with the usual standard cooling ($Q$) emissivities. For the DM candidate we focus in our work, $m_\chi=0.1$ GeV,  we find that it is possible to temporarily balance MURCA reactions giving rise to a temperature plateau. This will result in higher temperatures than expected for higher assumed NS masses. We have also considered the fact that LDM could act as a heat carrier involving an additional contribution to the SM conductivity, $\kappa_\chi$, allowing for a more efficient heat transport to the surface of the star at early times. In our simplified treatment and for our exploratory case study we have obtained trends that we expect could be robust against more refined models that are under investigation and will be reported elsewhere. Recently reported cold NSs can put additional constraints on the dozen-MeV range DM candidates in the very old ages. The realistic picture in the present scenario will be determined by the actual confirmation of existence of a dark matter particle sector and its nature and properties.
\vspace{6pt}

\acknowledgments{We acknowledge financial support from Junta de Castilla y Le\'on through the grant SA096P20, Agencia Estatal de Investigaci\'on through the grant PID2019-107778GB-100, Spanish Consolider MultiDark FPA2017-90566-REDC and PHAROS COST Actions MP1304 and CA16214. DB acknowledges support from the program Ayudas para Financiar la Contratación Predoctoral de Personal Investigador (Orden EDU/1508/2020) funded by Consejería de Educación de la Junta de Castilla y León and European Social Fund. We thank  M. Cermeño for help with parametrization in Eq. (\ref{qxx}) and comments, and D. Blaschke,  D.N. Voskresensky for useful comments.}






\end{document}